\documentclass{article}

\begin{document}

\rightline{September 2006}

\vskip 2cm

\centerline{\bf \huge Natural electroweak symmetry breaking}

\vspace{2mm}

\centerline{\bf \huge in generalised mirror matter models}

\vskip 2.2cm

\centerline{\large R. Foot and Raymond R. Volkas\footnote{ E-mail address:
rfoot@unimelb.edu.au, r.volkas@physics.unimelb.edu.au}}

\vspace{0.7cm}

\centerline{\large \it School of Physics}

\vspace{1mm}

\centerline{\large \it Research Centre for High Energy Physics}

\vspace{1mm}

\centerline{\large \it University of Melbourne,}

\vspace{1mm}

\centerline{\large \it Victoria 3010 Australia}

\vspace{1.2cm}

It has recently been pointed out that the mirror or twin Higgs model is
more technically natural than the standard model, thus alleviating the
``little'' hierarchy problem.  In this paper we generalise the analysis
to models with an arbitrary number of isomorphic standard model
sectors, and demonstrate that technical naturalness increases with
the number of additional sectors. We consider two kinds of models.
The first has $N$ standard model sectors symmetric under arbitrary
permutations thereof. The second has $p$ left-chiral standard model
sectors and $p$ right-chiral or mirror standard model sectors, with
$p$-fold permutation symmetries within both and a discrete
parity transformation interchanging left and right.
In both kinds of models the lightest scalar has an invisible
width fraction $1/N$, which will provide an important means
of experimentally testing this class of models.

\newpage

\section{Introduction}

One simple way to explain non-baryonic dark matter is to postulate 
the existence of a mirror sector (for an up-to-date review, see ref.\cite{review}). 
In this theory, each type of ordinary particle (other than the graviton)
has a distinct mirror partner. The ordinary and mirror particles form 
parallel sectors each with gauge symmetry $G_{SM} \equiv SU(3)_c \otimes SU(2) \otimes U(1)_Y$,
so that the overall gauge group is $G_{SM}\otimes G_{SM}$\cite{flv}. The interactions of each 
sector are governed by a Lagrangian of 
exactly the same form, except with left- and right-chiral fermions interchanged. In other words,
the Lagrangian has the form
\begin{eqnarray}
{\cal L} = {\cal L}_{SM} (e_L, e_R, q_L, q_R, W_{\mu}, B_{\mu},...) \ + \ 
{\cal L}_{SM} (e'_R, e'_L, q'_R, q'_L, W'_{\mu}, B'_{\mu},...) + {\cal L}_{mix}.
\end{eqnarray}
There are just two renormalisable gauge invariant interactions which
can couple the ordinary and mirror sectors together\cite{flv}:
\begin{eqnarray}
{\cal L}_{mix} = \epsilon F^{\mu \nu}F'_{\mu\nu} + 2\lambda \phi^{\dagger} \phi {\phi'}^{\dagger} \phi'
\end{eqnarray}
where $F^{\mu \nu} \equiv \partial^{\mu} B^{\nu} - \partial^{\nu} B^{\mu}$ [$F'^{\mu\nu} \equiv \partial^{\mu}B'^{\nu} - \partial^{\nu}B'^{\mu}$]
is the $U(1)$ [mirror $U(1)$] field strength tensor and $\phi, \phi'$ are the 
ordinary and mirror Higgs doublets. (If singlet neutrinos are added to both sectors,
then mass-mixing terms like $\overline{\nu}_R \nu'_L$ are also allowed in ${\cal L}_{mix}$.)

An interesting effect of the Higgs--mirror-Higgs mixing term (in ${\cal
L}_{mix}$) is to cause each 
of the two weak eigenstate Higgs fields ($\phi, \phi'$) to be maximal mixtures of the mass eigenstates, $h_+, h_-$.
Each of $h_{\pm}$ can be produced
 in colliders, but with production cross sections suppressed by 
a factor of $1/2$ compared to the standard model Higgs\cite{flv2}. Furthermore, each mass
 eigenstate will decay into the mirror sector half of the time -- giving another 
characteristic prediction of the theory\cite{flv2}. The Higgs-mirror Higgs coupling
 can also be reconciled with standard big bang nucleosynthesis in low reheat temperature 
scenarios\cite{ignatiev}.

Another interesting feature of the Higgs sector in 
these models is 
that it can\cite{zac,hall} alleviate the hierarchy problem 
because there is a limit in which the Higgs is, in part, a
pseudo-Goldstone boson\cite{georgi}.
This can most readily be understood if 
the Higgs potential is written in the form
\begin{eqnarray}
V = -\mu^2 (\phi^{\dagger} \phi + \phi'^{\dagger}\phi') + \lambda (\phi^{\dagger} \phi + 
\phi'^{\dagger} \phi')^2 + \delta [(\phi^{\dagger}\phi)^2 + (\phi'^{\dagger}\phi')^2].
\end{eqnarray}
The Higgs potential preserves a $U(4)$ symmetry in the limit
$\delta \to 0$, with the $\phi, \phi'$ transforming as the 4 representation of
$U(4)$. There are two non-trivial vacua, depending on 
whether $\delta < 0$ or $\delta > 0$. The symmetric vacuum occurs for $\delta > 0$ 
and this is the case to be considered in this paper\footnote{The mirror model with 
asymmetric vacuum ($\delta < 0$) has been studied in Ref.\cite{asym}.}.
In this case $\langle \phi \rangle = \langle \phi' \rangle \equiv u,$
with $u^2 = \mu^2/(4\lambda+2\delta)$.

Quadratically divergent corrections to the Higgs potential come from 1-loop
top quark, gauge boson and scalar Feynman diagrams. 
The top quark loop corrections have the form
\begin{eqnarray}
\mu^2 = \mu_0^2 + a_t \Lambda_t^2,
\end{eqnarray}
where $\mu^2_0$ is the bare parameter, $a_t = 3\lambda_t^2/8\pi^2$ and $\lambda_t \sim 1$ is the top
quark Yukawa coupling.  The parameter $\Lambda_t$ is the ultraviolet cutoff in the
naive cut-off regularisation approach.
The quadratic divergence in the mirror sector is of exactly the same form,
{\it so the quadratic divergences preserve
the $U(4)$ symmetry}. In the $\delta \to 0$ $U(4)$ symmetry limit, the spontaneous
breaking is $U(4) \to U(3)$. This implies seven Goldstone 
bosons, six of which are eaten by the $W^{\pm},Z$ and $W'^{\pm},Z'$, leading to one massless 
Higgs boson. In other words, in the $U(4)$ symmetry limit, one of the two 
physical scalars becomes massless. Of course, we do not expect $U(4)$ to 
be an exact symmetry of the potential: it is not a symmetry
 of the rest of the Lagrangian, and we know from experiments that
$m_{h_+}, \ m_{h_-} \stackrel{>}{\sim} 114 \ GeV$.
But it is an approximate symmetry when $\delta \stackrel{<}{\sim} \lambda$.

The hierarchy problem due to top quark loops is alleviated in the mirror
model, because the correction becomes
\begin{eqnarray}
{\delta \mu^2 (top) \over \mu^2} = {3\lambda_t^2 \over 4\pi^2}
{\Lambda_t^2 \over m_{h_+}^2},
\label{top2}
\end{eqnarray}
which is the same formula as in the standard model except with $m_{higgs} \to
m_{h_+}$.
In the standard model, the bound from
precision electroweak measurements is $m_{higgs} < M_{EW}$,
where $M_{EW} \approx 186$ GeV (which is the 95\% C.L. limit given
by the particle data group \cite{pdg}). 
However, in the mirror model, this bound 
becomes\cite{hall}
\begin{eqnarray}
m_{h_+}m_{h_-} &<& M_{EW}^2. 
\label{bound}
\end{eqnarray}
Evidently, a heavy $h_+$ can be compensated by a relatively light $h_-$. 
In fact the bound, Eq.(\ref{bound}), implies a limit of
$m_{h_+} \stackrel{<}{\sim} 300$ GeV (given that $m_{h_-}
\stackrel{>}{\sim} 114$ GeV).
Because of the larger $m_{h_+}$ limit, 
the fine tuning in the
$\mu^2$ parameter due to top quark loops is alleviated.  

Recently, the mirror matter model has been generalised to incorporate 
$N$ sectors\cite{f05}. 
The minimal standard model corresponds to $N=1$, the mirror model corresponds to $N=2$, but
in general there can be $N$ sets of particles. 
These generalised mirror models can also be motivated by the dark
matter problem and are therefore of significant interest.
In this general case there 
are $N$ physical scalars, one for each sector. How the Higgs physics
generalises in this $N$-sector case is an interesting 
question, and 
the purpose of this letter is to answer that question. We consider two
physically distinct, but related models. 
First, we consider having
the $N$ sectors exactly identical, so that a discrete $S_N$ (permutation 
symmetry of $N$ objects) is preserved. In this case there is no exact parity symmetry.
In the second case, an exact parity symmetry is required to exist, which
means that there are $p$ ordinary isomorphic sectors and $p$ isomorphic mirror 
sectors (so that $N=2p$ is necessarily even in this case).
The ordinary and mirror sectors are related to each other by interchanging the left- and
right-handed chiral fermions, but are otherwise identical. Both types of models 
alleviate the hierarchy problem in a similar way.

\section{The SM generalised to $N$ isomorphic sectors}

The SM generalised to $N$ isomorphic sectors is described by the Lagrangian
\begin{eqnarray}
{\cal L} = \sum^{N}_{i=1} {\cal L}_{SM} (e_{iL}, e_{iR}, q_{iL}, q_{iR}, W_i^{\mu}, B_i^{\mu},...)
+ {\cal L}_{mix},
\end{eqnarray}
where we use the integer subscripts to label the particles 
from the $N$ sectors. Clearly the Lagrangian has gauge symmetry 
$G_{SM}^{N}$ and discrete symmetry $S_{N}$.
The ${\cal L}_{mix}$ part 
describes the interactions coupling ordinary and mirror particles together
which are consistent with these symmetries.
In general, ${\cal L}_{mix}$ has the form
\begin{eqnarray}
{\cal L}_{mix} = \epsilon \sum^{N}_{k,l=1} F_{k}^{\mu \nu} F_{l \mu \nu} + 
2 \lambda \sum^{N}_{k,l=1} \phi^{\dagger}_k \phi_k \phi^{\dagger}_l \phi_l
\end{eqnarray}
where $k \neq l$ in the sums and $F^{\mu \nu}_{i} \equiv \partial^{\mu} B_i^{\nu} - \partial^{\nu} B_i^{\mu}$.

The most general Higgs potential can be expressed as
\begin{eqnarray}
V = -\mu^2 \sum^{N}_{i=1} \phi^{\dagger}_i \phi_i
\ + \ \lambda \left[ \sum^N_{i=1} \phi^{\dagger}_i \phi_i \right]^2
\ + \ \delta  \sum^N_{i=1} (\phi^{\dagger}_i \phi_i)^2 \ . 
\label{v3}
\end{eqnarray}
In the limit $\delta \to 0$, the potential exhibits a $U(2N)$ symmetry.
For $\delta > 0$, the minimum of this potential occurs 
when each $\langle \phi_i \rangle = u$ where
\begin{eqnarray}
u^2 = {\mu^2 \over 2N\lambda + 2\delta} \ .
\end{eqnarray}
The parameters are chosen so that $u \simeq 174$ GeV.
The discrete symmetry $S_{N}$ is not spontaneously broken; it
is an exact symmetry of both the vacuum and the Lagrangian\footnote{The
alternative cases where the discrete symmetry is broken, 
either spontaneously or explicitly (by soft breaking terms in the 
the Higgs potential\cite{hall}) are interesting, but beyond the
scope of the present work.}.

In each sector, three of the scalar degrees of freedom are `eaten' 
by the $W^{\pm}, Z$ gauge
bosons from that sector, leaving 
one physical scalar per sector. Thus
there are $N$ physical scalar bosons, which we denote by $h_i$ $(i = 1,...,N)$. 
The mass matrix for these physical scalar bosons can be obtained from the 
Higgs potential, Eq.(\ref{v3}), by expanding around the vacuum, and is
\begin{eqnarray}
M^2 = \left(\begin{array}{ccccccc}
x+y & x & x & . & . & x & x \\
x & x+y & x & . & . & . & x \\
x & x & x+y & x & . & . & x \\
. & . & x & . & . & . & . \\
. & . & . & . & . & x & . \\
x & . & . & . & x & x+y & x \\
x & x & . & . & x & x & x+y 
\end{array}\right),
\end{eqnarray}
where $x \equiv 4\lambda u^2$, $y \equiv 4\delta u^2$.
The matrix has characteristic equation
\begin{eqnarray}
det(M^2 - \stackrel{\sim}{\lambda} I) 
= (\stackrel{\sim}{\lambda} - y )^{N-1} (\stackrel{\sim}{\lambda} - Nx - y) = 0.
\end{eqnarray}
Thus, $N-1$ of the scalars ($h_2,...,h_N$) are degenerate, 
with $m^2 = 4\delta u^2$,
and there is one (heavier) scalar, $h_1$, with mass 
$m^2_{h_1} = 4(N\lambda + \delta) u^2 = 2\mu^2$.

The $N\times N$ orthogonal transformation matrix $O$ relating the weak eigenstates to the mass eigenstates
is most usefully written in the form
\begin{eqnarray}
O = {1 \over \sqrt{N}} \ \left(\begin{array}{ccccccc}
1 & \sqrt{N-1} & 0 & 0 & . & . & 0 \\
1 & -\epsilon_1 & (N-2)\epsilon_2 & 0 & . & . & 0 \\
1 & -\epsilon_1 & -\epsilon_2 & (N-3)\epsilon_3 
 & . & . & . \\
1 & -\epsilon_1 & -\epsilon_2 & -\epsilon_3 & . & . & . \\
. & . & . & . & . & . & 0 \\
. & . & . & . & . & . & \epsilon_{N-1} \\
1 & -\epsilon_1 & -\epsilon_2 & -\epsilon_3 & . & . & -\epsilon_{N-1} \\
\end{array}\right)
\end{eqnarray}
where 
\begin{equation}
\epsilon_i^2 = \frac{N}{(N-i)(N+1-i)}.
\end{equation}
There is no unique choice for $O$
since $N-1$ of the scalars are degenerate.  The above equation refers to one possible
basis.

In this basis, the weak eigenstate scalar $\phi_1$, coupling to 
the particles of the first sector
(which we will choose to be the standard particles), is 
a superposition of just two mass eigenstates,
\begin{eqnarray}
\phi_1 = {1 \over \sqrt{N}} \ h_1 + \sqrt{{N-1 \over N}} \ h_2,
\end{eqnarray}
where $h_1$ is the heavier state.
Evidently, the $h_2$ state couples to the standard fermions and gauge bosons
just like the standard model Higgs, except with coupling reduced by a 
factor $\sqrt{(N-1)/N}$. 
The heavier state, on the other hand,
couples to the standard particles with a coupling reduced by a factor of $1/\sqrt{N}$.
(For the mirror model case of $N=2$, both factors reduce to
$1/\sqrt{2}$ obtained in ref.\cite{flv2}).
Each of these scalars also couples to the particles of the other sectors,
which means their invisible width contribution will 
be $1/N$ for the lighter scalar and $(N-1)/N$ for the heavy scalar.
The decays of $h_3,\ldots,h_N$ are entirely invisible.
This characteristic prediction of these models can be probed in
forthcoming collider
experiments such as the LHC. This will supply an important test
of these models, provided that $N$ is not too large. It is also 
interesting to note that in the large $N$ limit, the interactions
of the lightest scalar reduce to those of the standard model Higgs.

Importantly, the quadratic divergences to the Higgs potential preserve the 
$U(2N)$ symmetry (which contains the gauged $SU(2)^N \otimes U(1)^N$ symmetry as 
a subgroup). Thus, in this $U(2N)$ symmetry limit ($\delta \to 0$), the spontaneous symmetry breaking 
pattern is $U(2N) \to U(2N-1)$, leading to $(2N)^2 - (2N-1)^2 = 4N-1$ Goldstone bosons,
of which $3N$ are eaten by the gauge bosons. This implies that $N-1$ of the $N$ scalars
 do not gain any mass from the quadratic corrections in the $U(2N)$ limit. 
Of course, the $U(2N)$ symmetry is explicity broken when $\delta \neq 0$ in the Higgs potential.
Thus one can have the $N-1$ degenerate scalars naturally light (e.g. $\stackrel{<}{\sim} 150\ GeV$) while the 
remaining scalar can be quite heavy (e.g.\ $\sim TeV$). 
In this way the (little) hierarchy problem can be alleviated. 
There is no conflict with 
the precision electroweak data, which prefer a light scalar, 
because the weak eigenstate
$\phi_1$ is composed mainly of the light state $h_2$, with only a small fraction of 
amplitude $1/\sqrt{N}$ of the heavy state.

Explicitly, in the standard model, the relevant radiative corrections for the 
electroweak precision  tests
involve $\log \ m_h$, from which
the bound $m_h < M_{EW} \approx 186\ GeV$ arises. 
In this model, 
\begin{eqnarray}
\log \ m_h \to {1\over N}\log \ m_{h_1} + {N-1 \over N}\log \ m_{h_2}. 
\end{eqnarray}
Thus the standard model bound, $m_h < M_{EW}$, is replaced by
\begin{eqnarray}
m_{h_1} m^{N-1}_{h_2}  < M_{EW}^{N}
\end{eqnarray}
This bound, for the special case of $N=2$, was obtained in
Ref.\cite{hall}.
Clearly, we can have one heavy state, $h_1$, with mass much greater than 
$M_{EW}$,
so long as the other state, $h_2$ is lighter than this bound. 
For the minimal mirror model case 
of $N = 2$, we can have $m_{h_1} \approx 300\ GeV$ for $m_{h_2}$ at the experimental
limit of $\approx 114\ GeV$. For increasing $N$, the limit on $m_{h_1}$ rapidly weakens,
allowing for TeV scale $m_{h_{1}}$ for $N \stackrel{>}{\sim} 4$.

The quadratically divergent corrections due to the top quark loops 
have the form
\begin{eqnarray}
{\delta \mu^2 (top) \over \mu^2} = {3\lambda_t^2 \over 4\pi^2}
{\Lambda_t^2 \over m_{h_1}^2},
\label{topq}
\end{eqnarray}
which will suppressed by having $m_{h_1}$ large.

Let us now consider the quadratic divergences due to scalar loops (the 
quadratic divergences due to gauge boson loops have the same form 
as the scalar loops, only they are smaller in magnitude). The quadratic correction from 
1-loop Higgs self-energy diagrams is
\begin{eqnarray}
\mu^2 \to \mu_0^2 - a_H \Lambda^2
\end{eqnarray}
where $\Lambda_H$ is the ultraviolet cutoff.
In the standard model, $a_H = 3\lambda/8\pi^2$ where $\lambda$
is the Higgs potential quartic coupling constant. With $N$ isomorphic 
sectors, this generalises to 
\begin{eqnarray}
a_H = {3\lambda + 2(N - 1)\lambda + 3\delta \over 8\pi^2} \ .
\end{eqnarray}
The corrections are qualitatively different from the top quark  
loops
in that they contain the factor $N$ (as well as 
being different in sign). 
These corrections can be put into the form
\begin{eqnarray}
{\delta \mu^2 (scalars) \over \mu^2} =
{-\Lambda_H^2 \over 4\pi^2 u^2} \left[
{2N+1 \over 4N}\left( 1 - \gamma\right) + {3\over 4} \gamma \right],
\label{scaler}
\end{eqnarray}
where $\gamma \equiv m^2_{h_2}/m^2_{h_1}$.
Evidently, the magnitude of these corrections is not greatly suppressed
for increasing $N$.
In fact, for  $N \stackrel{>}{\sim} 2$
these corrections are approximately independent of
$N$, and taking $\gamma \ll 1$, we find
\begin{eqnarray}
{\delta \mu^2 (scalars) \over \mu^2} = 
 - { \Lambda^2_H \over 8\pi^2 u^2}.
\end{eqnarray}
Assuming no more than $10\%$ fine tuning $|\delta \mu^2 (scalars)/\mu^2|
\stackrel{<}{\sim} 10$, we obtain an upper limit on $\Lambda_H$ of
\begin{eqnarray}
\Lambda_H \stackrel{<}{\sim} 5 \ {\rm TeV}
\end{eqnarray}
For large $m_{h_1}$ this correction dominates over the top
quark correction (assuming identical cutoffs, $\Lambda_t = \Lambda_H$), 
and this becomes the scale for new physics.
This is a significant improvement over the standard model,
since the electroweak precision measurements imply
$m_{higgs} < M_{EW} \simeq 186\ GeV$, and, 
in that domain, the top quark loop dominates the corrections to
$\mu^2$. Requiring  $|\delta \mu^2 (top)/\mu^2| < 10$ 
gives $\Lambda_t \stackrel{<}{\sim} 2$ TeV.

\section{Mirror Higgs models with $p$ ordinary and $p$ mirror sectors}

We now consider the mirror matter case, which requires that there are $p$ ordinary and
$p$ mirror sectors, a total of $N=2p$ sectors.
The Lagrangian describing this case has the form
\begin{eqnarray}
{\cal L} &=& \sum^p_{i=1} {\cal L}_{SM} (e_{iL}, e_{iR}, q_{iL}, q_{iR}, W^\mu_i, B_i^\mu, ...) 
\nonumber \\
\ &+& \ \sum^p_{i=1} {\cal L}_{SM} ({e}'_{iR}, {e}'_{iL}, {q}'_{iR},
{q}'_{iL}, {W'}^\mu_i, {B'}_i^\mu, ...) \ + \ 
{\cal L}_{mix},
\end{eqnarray}
where we use the integer subscripts to label the particles from the $p$ ordinary sectors and
primes plus integer subscripts to label their corresponding mirror partners. In this generalised mirror parity
symmetric case, ${\cal L}_{mix}$ has the form\cite{f05}
\begin{eqnarray}
{\cal L}_{mix} &=& \epsilon \sum^p_{i=1} F_i^{\mu \nu} \sum^p_{j=1} F^{'}_{j\mu \nu} \ + \ 
\epsilon' \sum^p_{k \neq l=1} (F_k^{\mu \nu} F_{l \mu \nu} + F_k^{'\mu\nu} F^{'}_{l\mu\nu})\nonumber \\
\ &+& \ (2\lambda+\delta_2) \sum^p_{i=1} \phi_i^{\dagger}\phi_i \sum^p_{j=1} {\phi'}^{\dagger}_j {\phi'}_j
\ + \ 2\lambda \sum^p_{k \neq l=1} (\phi_k^{\dagger}\phi_k \phi_l^{\dagger} \phi_l
+ {\phi'}^{\dagger}_k {\phi'}_k {\phi'}_l^{\dagger} {\phi'}_l) \ , 
\end{eqnarray}
where $F_i^{\mu \nu} \equiv \partial^\mu B_i^\nu - \partial^\nu B_i^\mu \
[F_i^{'\mu \nu} \equiv \partial^\mu B_i^{'\nu} - \partial^\nu B_i^{'\mu}]$.
Note that the second and fourth terms only exist for $p \ge 2$.

For the special case of $p=1$, the Higgs physics of this model is the same as 
the corresponding case of having two exactly isomorophic sectors.
This is because the discrete symmetry in both cases is the same: $Z_2$.
However, for $N \ge 4$ the discrete symmetry is quite distinct:
$S_N$ (for the case of $N$ isomorphic sectors) versus $Z_2 \otimes S_p \otimes S_p$ (for the mirror parity case).
In fact, the latter case has an extra parameter (for $p \ge 2$) in 
the Higgs potential:
\begin{eqnarray}
V &=& -\mu^2 \left[\sum^p_{i=1} \phi_i^{\dagger} \phi_i + 
{\phi'}_i^{\dagger}\phi'_i
\right]\ + \
\lambda\left[ \sum^p_{i=1} \phi_i^{\dagger} \phi_i + {\phi'}_i^{\dagger}\phi'_i
\right]^2 \nonumber \\ \ &+& \
\delta_1 \left[ \sum^p_{i=1} (\phi^{\dagger}_i \phi_i)^2 + ({\phi'}^{\dagger}_i \phi'_i)^2
\right] \ + \
\delta_2 \left[ \sum^p_{i=1} \phi^{\dagger}_i \phi_i \right] 
\left[ \sum^p_{i=1} {\phi'}^{\dagger}_i \phi'_i \right] \ .
\label{v5}
\end{eqnarray}
For $\delta_1 > 0$ and $2\delta_1 - p\delta_2 > 0$, the potential is 
minimised when
each of the $\phi_i, \phi'_i$ gain identical VEVs, $u$, where
\begin{eqnarray}
u^2 = {\mu^2 \over 2\lambda N + 2\delta_1 + p\delta_2} \ .
\end{eqnarray}

As before, in each sector, three of the scalar degrees of freedom are
`eaten' by the $W^{\pm}, Z$ bosons from that sector, leaving 
one physical scalar per sector. Thus
there are $N$ physical scalar bosons, which we denote by $\phi_i, \phi'_i$, $(i = 1,...,p)$. 
The mass matrix for these physical scalar bosons can be obtained from the above 
Higgs potential, Eq.(\ref{v5}), by expanding around the vacuum, and is
\begin{eqnarray}
M^2 = \left(\begin{array}{cc}
M_1^2 & M_2^2 \\
M_2^2 & M_1^2
\end{array}\right),
\end{eqnarray}
where
\begin{eqnarray}
M^2_1 = \left(\begin{array}{cccccc}
x+y & x & . & . & . & x \\
x & x+y & x & . & . & x \\
x & x & x+y & x & . & . \\
. & . & x & x+y & x & . \\
. & . & . & x & . & . \\
x & . & . & . & . & x+y 
\end{array}\right),\  \ 
M^2_2 = \left(\begin{array}{cccccc}
x+z & x+z & . & . & . & x+z \\
x+z & x+z & . & . & . & . \\
. & . & . & . & . & . \\
. & . & . & . & . & . \\
. & . & . & . & . & . \\
x+z & x+z & . & . & . & x+z 
\end{array}\right)
\end{eqnarray}
and $x = 4\lambda u^2$, $y = 4\delta_1 u^2$, $z = 2\delta_2 u^2$.
The matrix has the characteristic equation
\begin{eqnarray}
det (M^2 - \stackrel{\sim}{\lambda} I) = (\stackrel{\sim}{\lambda} - y )^{2p-2} 
[\stackrel{\sim}{\lambda} - y + pz][\stackrel{\sim}{\lambda} - y - pz - 2px] = 0.
\end{eqnarray}
Thus, we have $2p-2$ degenerate mass eigenstate scalars ($h_3,...,h_N$) with 
$m^2 = 4\delta_1 u^2$,
and  two states, $h_1, \ h_2$ with $m^2_{h_1} = 2(2\delta_1 + 4p \lambda
+ p\delta_2)u^2 = 2\mu^2$ and $m^2_{h_2} = 2(2\delta_1 - p\delta_2)u^2$, respectively.
The $N\times N$ orthogonal matrix transformation, relating the weak eigenstates to the mass eigenstates,
can be written in the following form, without loss
of generality:
\begin{eqnarray}
O = {1 \over \sqrt{N}}\left( \begin{array}{cccccccccccccc}
1 & 1 & | & \sqrt{N-2} & 0 & . & . & 0 & | & 0 & 0 & . & . & 0  \\
1 & 1 & | & -\epsilon_1 & (p-2)\epsilon_2 & 0 & . & . & | & 0 & . & . & . & 0 \\
. & . & | &  . & -\epsilon_2 & . & . & . & | & . & . & . & . & . \\
. & . & | &  . & . & . & . & 0 & | & . & . & . & . & . \\
1 & 1 & | &  -\epsilon_1 & -\epsilon_2 & . & . & \epsilon_{p-1} & | & 0 & . & . & . & 0\\
1 & 1 & | & -\epsilon_1 & -\epsilon_2 & . & . & -\epsilon_{p-1} & | & 0 & 0 & 0 & 0 & 0 \\
- & - & - & - & - & - & - & - & - & - & - & - & - & - \\
1 & -1 & | & 0 & 0 & . & . & 0 & | & \sqrt{N-2} & 0 & . & . & 0    \\
1 & -1 & | & 0 & . & .  & . & 0 & | & -\epsilon_1 & (p-2)\epsilon_2  & 0 & . & .  \\
. & . & | &  0 & . & . & . & 0 & | & . & -\epsilon_2 & . & . & . \\
. & . & | &  0 & . & . & . & 0 & | & . & . & . & . & 0 \\
1 & -1 & | &  0 & . & . & . & 0 & | & -\epsilon_1 & -\epsilon_2 & . & . & \epsilon_{p-1} \\
1 & -1 & | &  0 & 0 & . & . & 0 & | & -\epsilon_1 & -\epsilon_2 & . & . & -\epsilon_{p-1} 
\end{array}\right) 
\label{u2}
\end{eqnarray}
where 
\begin{eqnarray}
\epsilon_i^2 \equiv {2p \over (p-i)(p+1-i)} \ .
\end{eqnarray}

In this basis, the weak eigenstate scalar, $\phi_1$, coupling to the
particles of the first sector
(which we will choose to be the standard particles), is composed of just three mass eigenstates
(for $p \ge 2$):
\begin{eqnarray}
\phi_1 = {1 \over \sqrt{N}} \ h_1 + {1 \over \sqrt{N}} \ h_2 
+ \sqrt{{N-2 \over N}} \ h_3 \ .
\end{eqnarray}
Evidently, the $h_3$ state couples to the standard fermions and gauge bosons
just like the standard model Higgs, except with coupling reduced by the 
factor $\sqrt{(N-2)/N}$. 
The two other states, $h_1,\ h_2$, on the other hand,
couple to the standard particles each with couplings reduced by 
factors of $1/\sqrt{N}$.
Each of these scalars also couples to the particles of the other sectors,
which means that the invisible width contribution from these particles will 
be $2/N$ for the lighter scalar and $(N-1)/N$ for the two heavier scalars.

Note that the $U(2N)$ symmetry limit corresponds to $\delta_1, \delta_2 \to 0$ and
in this limit only one of the scalars, $h_1$, has a mass. Clearly the naturalness results
obtained for the case of the standard model generalised to $N$
isomorphic sectors (considered in section 2) carry through to the
above generalised mirror parity symmetric model.
In particular, Eq.(\ref{topq}) has exactly the same form, while
Eq.(\ref{scaler}) becomes:
\begin{eqnarray}
{\delta \mu^2 (scalars) \over \mu^2} =
{-\Lambda_H^2 \over 4\pi^2 u^2} \left[
{2N+1 \over 4N} \ + \ {1 \over 4N} \gamma_1 \ + \ \left({1 \over 4} - {1
\over 2N}\right)\gamma_2 \right],
\label{scaler2}
\end{eqnarray}
where $\gamma_1 \equiv m^2_{h_2}/m^2_{h_1}$ and $\gamma_2 \equiv
m^2_{h_3}/m^2_{h_1}$.
The bound from precision electroweak measurements generalizes to:
\begin{eqnarray}
m_{h_1} m_{h_2} m_{h_3}^{N-2} < M^N_{EW} .
\end{eqnarray}
Thus, as in the model of the previous
section, $m_{h_1}$ can be large ($\stackrel{>}{\sim}$ TeV) for
$N \stackrel{>}{\sim} 4$, if $m_{h_2}, m_{h_3}$ are light ($\sim 120$
GeV). In this large $m_{h_1}$ regime the quadratic corrections to
$\mu^2$ are dominated by $\delta\mu^2 (scalars)$, giving the same limit of
$\Lambda \stackrel{<}{\sim} 5$ TeV, as we found for the previous
model.

\section{Conclusion}

In conclusion, we have examined naturalness of the electroweak symmetry breaking
 in generalised mirror matter models. We considered two cases: a) where there are $N$ isomorphic standard model 
sectors completely symmetric under the permutation symmetry $S_N$, and 
b) where there are $p$ ordinary and $p$ mirror sectors 
(giving a total of $N=2p$ sectors) completely
symmetric under ${\cal P} \times S_p \times S_p$, where ${\cal P}$ is the mirror
parity symmetry. Previous work has shown that the $N=2$ case alleviates the hierarchy problem 
and we have shown that increasing $N$ further reduces the need for fine tuning.
The end result is that such models can naturally accommodate a 5 TeV scale 
cut-off consistently with precision electroweak measurements.
In other words, these models can alleviate the little hierarchy problem.
Furthermore, the models 
are extremely simple, with only a few parameters beyond 
the minimal standard model, and offer one interesting direction for new
physics at the TeV scale.

\vskip 1cm

\noindent
{\bf Acknowledgements:}
This work was supported by the Australian Research Council.


\begin{thebibliography}{999}


\bibitem{review}
R. Foot, Int. J. Mod. Phys. D13, 2161 (2004) [astro-ph/0407623].

\bibitem{flv}
R. Foot, H. Lew and R. R. Volkas, Phys. Lett. B272, 67 (1991).
The idea that an exact unbroken parity symmetry can be
respected by the fundamental interactions if a set
of mirror particles exist was discussed, prior to the advent
of the standard model of particle physics in:
T. D. Lee and C. N. Yang, Phys. Rev. 104, 256 (1956) and
further developed in I. Kobzarev, L. Okun, I. Pomeranchuk, Sov. J. Nucl. Phys. 3, 837 (1966).

\bibitem{flv2}
R. Foot, H. Lew and R. R. Volkas, Mod. Phys. Lett. A7, 2567 (1992).


\bibitem{ignatiev}
A. Yu. Ignatiev and R. R. Volkas, Phys. Lett. B487, 294 (2000) [hep-ph/0005238].

\bibitem{zac}
Z. Chacko, H-S. Goh and R. Harnik, Phys. Rev. Lett. 96, 231802 (2006)
[hep-ph/0506256].


\bibitem{hall}
R. Barbieri, T. Gregoire and L. J. Hall,
hep-ph/0509242.

\bibitem{georgi}
H. Georgi and A. Pais, Phys. Rev. D10, 539 (1974);
Phys. Rev. D12, 508 (1975); D. B. Kaplan and H. Georgi,
Phys. Lett. B136, 183 (1984); D. B. Kaplan, H. Georgi and 
S. Dimopoulos, Phys. Lett. B136, 187 (1984);
H. Georgi and D. B. Kaplan, Phys. Lett. B145, 216 (1984);
N. Arkani-Hamed, A. G. Cohen and H. Georgi,
Phys. Lett. B513, 232 (2001).

\bibitem{asym}
R. Foot, H. Lew and R. R. Volkas, JHEP 0007, 032 (2000) [hep-ph/0006027];
R. Foot and H. Lew, hep-ph/9411390.

\bibitem{pdg}
W.-M. Yao {\it et al.}, J. Phys. G 33, 1 (2006).


\bibitem{f05}
R. Foot, Phys. Lett. B632, 467 (2006) [hep-ph/0507294].

\end{thebibliography}
\end{document}